# Mendelian randomization and causal networks for systematic analysis of omics

Azam Yazdani, Ph.D.

Brigham Women's Hospital, Harvard Medical School

ayazdani@bwh.harvard.edu

**Abstract.** Mendelian randomization implemented through instrumental variable analysis is frequently discussed in causality and recently the number of applications on real data is increasing. However, there are very few discussions to address modern biomedical questions such as the integration of large scale omics in causality. While in the age of large omics, we face several hundred or thousands of components with little knowledge about the underlying structures, the focus of the field is on small scales and mostly with known structures. The availability of large omic data accentuates the need for techniques to identify interconnectivity among the omic's components and reveal the principles that govern the relationships.

This study extends instrumental variable techniques to identify causal networks in large scales and assess the assumptions. Large-scale causal networks are complex and further analyses are required to uncover mechanisms by which the components are related within and between omics and linked to disease endpoints. This study will review these utilities of causal networks for mechanistic understanding.

**Introduction**. To understand disease mechanisms, one must first characterize the topology and dynamics of relationships among individual components of the physiological system (e.g., metabolites, proteins, gene expressions) (1). Rapid advances in network biology indicate that biological processes are influenced by a common set of rules that could potentially revolutionize our view of biology and disease pathology (2). Identification of interconnectivity among observed components as networks provides fundamental insights into complex biological processes that would not be revealed by focusing on individual biological units of the network (3).

Causal networks are ideally suited for analyzing multi-omic and heterogeneous data sets to establish the architecture for a cell or tissue, thereby providing a mechanistic understanding of cellular processes and identifying ways to intervene upon them (4). Using the principles of Mendelian randomization on a genome-wide scale and then integrating genetic with other omics allows us to relate information from different omics in a cohesive analytic framework and to identify causal networks. The number of applications for causal inference is growing recently (5)(6)(7)(8)(9). However, the focus of applications is on small scale in terms of the number of components under consideration and mostly with the knowledge about the underlying structure (a major source is (10) where the relationship/structure of the exposure, response and covariates are known; therefore, we could identify mediators and confounders). But, in omics, we face with several hundreds or thousands of components with little knowledge about the interconnectivity. Therefore, development of analytical approaches that systematically analyze large-scale data in a causal framework is essential to extract the meaning from the data and to elucidate the complexity of the system under study.

Here, we extend instrumental variable (IV) techniques for identification of causal networks in large scale and explain the assessment of IV assumptions. Large-scale causal networks are complex and cannot be informative without further analysis. Thus, we describe some utilities of the causal networks to uncover the underlying processes for understanding the system under study.

**An overview of instrumental variable approach**. The IV approach has been discussed in multiple publications, e.g., (11)(12)(13)(14)(15)(16). Here, we discuss some of the points for further clarification. It is known that causal inference could not be achieved through a pure analysis of observational data and some knowledge is required. Here, the knowledge that makes causal inference possible is that genetic inherited variation is a cause of phenotypic variation and not the other way around.

The idea behind application of IV is using variation in the system that is free of confounding to assess causal inference in observational studies as if we have randomization. Therefore, it is natural to use genetically randomized effects as IVs that are less susceptible to confounding by hidden variables (17)(18). Similar to randomization that the effect of random assignment of treatment reaches the response only through the exposure/treatment, the effect of an IV on the response is only through the exposure; Figure 1 illustrates this assumption. If the assumption is not satisfied, the genetic variant is not qualified as an IV and the genetic variant itself confounds the study. In Figure 1, the question mark on the arrow connecting the exposure and response represents the question under study, which is the aim in IV applications. Missing arrows between IV and the response as well as the IV and the confounder mean the effect of IV to the response is only through the exposure. Therefore, genetic variants with pleiotropic effects on both exposure and response or genetic variants associated with (un)observed confounders are not qualified as IVs since they open paths to the response which are not through the exposure; and as a result, we could not simulate randomization. Meanwhile, IV must have a strong relationship with the exposure to predict variation in it significantly, thereby producing stability of results (14).

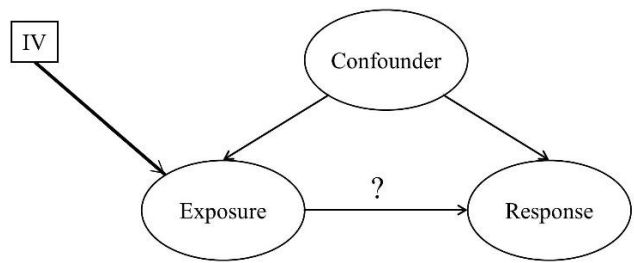

**Figure 1**. Illustration of the question and assumptions in IV technique. The question mark represents the purpose in IV applications. The missing arrows and links represent the assumptions. The thick link between the IV and exposure represents the strong relationship between these two.

**Large-scale causal networks**. The goal of finding causal relationships in large scale studies, i.e., causal networks, is to analyze a system comprehensively and uncover underlying biochemical networks for mechanistic understanding. Causal networks are Bayesian networks augmented with Mendelian randomization/IV techniques. In Figure 1, the IV is depicted in a different shape, a rectangular, to represent that IV is from a different granularity and is used as a tool to identify causal relationship between an exposure and a response. Similarly, in large-scale causal networks, IVs are used to identify causal relationships among a large number of components, i.e., variables of interest such as traits, risk factors, or molecular status. Note that in identification of causal networks, each variable can be an exposure for another variable and a response for a third variable at the same time.

IV techniques are for assessing <u>causal</u> relationship and not finding the relationship between variables of interest. This leads us to the principles below.

**Principle 1.** Dependency between variables of interest cannot vanish given any genetic IV.

**Principle 2.** Dependency between a variable of interest, e.g., a response, and a genetic IV can vanish given another variable of interest, e.g., an exposure.

As explained in the previous section, to assess causal relationship between two observed variables, i.e., one exposure and a response, one IV is required. Consider $Y$ and $Z$ as two random variables representing two variables of interest and $G$ as a random variable representing genetic information:

If $Y \perp Z | G$, then the causal network is $Y \leftarrow G \rightarrow Z$.

If $Y \perp G | Z$, then the causal network is $G \rightarrow Z \rightarrow Y$.

If $Z \perp G | Y$, then the causal network is $G \rightarrow Y \rightarrow Z$.

For identification of causal relationships among three variables, at least two IVs are required. However, the two IVs could be dependent. Consider three random variables $Z_1, Z_2$ and $Z_3$ and two dependent genetic IVs $X_1$ and $X_2$ as below. Any of the conditional (in)dependency below leads us to the causal network depicted under it:

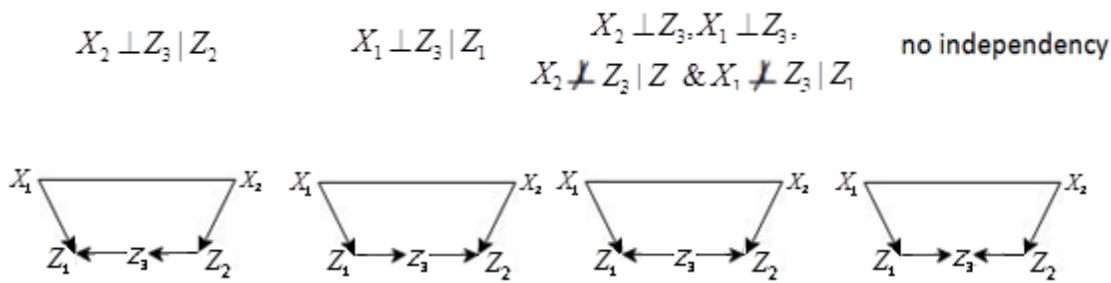

Note that when we call a variable an IV that means the IV assumption is satisfied.

For identification of causal relationships among four variables, two dependent IVs is not sufficient and a third IV is required and so on. Although the genome includes millions of SNPs, it is difficult to find appropriate IVs to identify a causal network in large scale. For example, Inouye et al. (19) had ~160 components (metabolites) in an analysis; using SNPs as IVs, they identified a causal network that included ~29 metabolites but had to exclude the others from the causal properties analysis due to the lack of SNPs strongly associated with the excluded metabolites. In addition, the properties of genetic variants, such as linkage disequilibrium structure, pleiotropic effects and population structure may violate IV assumptions, especially in large-scale omics data that include highly correlated components and genetic variants may have pleiotropic effects on the components. Therefore, the challenge is finding a sufficient number of IVs that satisfy assumptions for identification of a large-scale causal network.

One way to address this challenge is instead of using natural IVs, such as genetic variants, generating IVs to construct causal networks. One approach for generating IVs is extracting information from genetic variants using statistical techniques, such as principal component analysis, see (20)(21). Since genetic variants are abundant, we can generate multiple instrumental variables sufficient for identification of causal networks in large scales. Obtaining a principal component with correlated SNPs generates a measure that explains a large amount of variation and thus generates a stronger association with the exposure. Since the generated IVs are independent, we do not need to be concerned about over fitting by allocating multiple IVs to each exposure to explain even a larger amount of variation in the exposure, see Figure 2, and therefore, produce stability in the causal network.

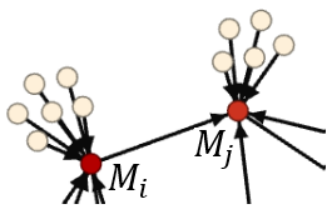

**Figure 2.** A subset of a large scale causal network. Red nodes are the components of an omic under study. The white nodes are genetic IVs. The IVs are not connected which is due to generating independent IVs using an unsupervised principal component analysis. Multiple IVs are allocated to each component to explain a higher portion of variation in each component. Using the IVs, the direction of effect between the components is identified which has causal interpretation ($M_i \rightarrow M_j$ means $M_i$ is a cause of $M_j$).

To satisfy the IV assumption, we select those IVs that satisfy the property below: for investigating the causal effect of $M_i$ on $M_j$ ($M_i \rightarrow M_j$), we assess

$$IV \perp M_j \mid M_i. \qquad (1)$$

This property means that the effect of an IV on the response (here, $M_j$) is through the exposure (here, $M_i$) and not the other variables/paths. If this property is not satisfied, the variable is not qualified as an IV to investigate $M_i \rightarrow M_j$. This property prevents violation of IV assumption.

**Utilities of causal networks**. Before reviewing some utilities of causal networks, we review differences between networks, Bayesian networks, and causal networks. Some networks are constructed on the basis of interactions, such as protein-protein interaction networks, or on the basis of pairs of genes with similar expression pattern across samples, such as co-expression networks. Bayesian networks are based on simultaneous analysis of all variables under investigation. Bayesian networks represent conditional (in)dependence properties and any relationship between two variables is after conditioning on the remaining variables in the study.

Consider the case with three variables in the system, A, B, and C. If each of these two variables are correlated, the visualization is the network in Figure 3.a, while the Bayesian network of these three variables can be Figure 3.b, which represents A and C are connected through B. From Figure 3, we can see how much Bayesian networks are more informative and sparser compared to networks especially in large-scale omics data that include a number of measured components that are highly correlated.

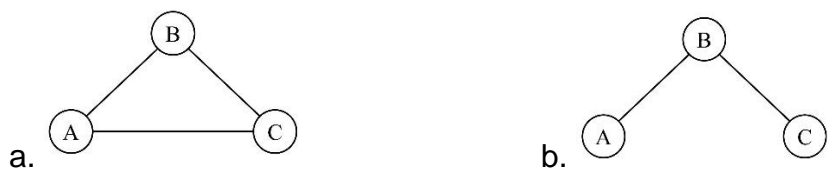

**Figure 3**. **a.** A network over variables of A, B, and C (visualization of pairwise correlations). **b.** A Bayesian network over variables of A, B, and C which is based on joint probability distribution of these three variables and a representation of conditional (in)dependence properties.

Causal networks are Bayesian networks augmented with principles of Mendelian randomization. Causal networks are robustly directed Bayesian networks and are more informative compared to Bayesian networks. While Bayesian networks provide information about conditional (in)dependence and reveal highly connected nodes, causal networks identify the role of each node individually (e.g. Figure 4) and as a group (i.e., module) in the system and principles governing the relationships (e.g. Figure 5) and therefore, facilitate mechanistic understanding of the system under study. For details see "Causal network parameters" in Appendix.

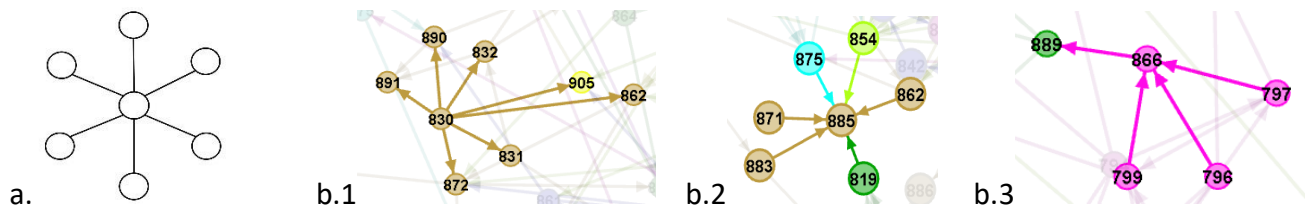

**Figure 4**. **a**. Extracted information from a Bayesian network. We can see a node highly connected with other nodes, a hub. **b**. Extracted information from a causal network about the role of a hub. A highly connected node/hub can be highly influential in the system (node 830 in b.1), or highly influenced by the system (node 885 in b.2), or a combination of both (node 866 in b.3). In b.1, effect of node 830 spreads in the system through the 7 nodes influenced directly by it. In b.2, effects of the 6 nodes around are blocked in node 885. In b.3, effects of the three pink nodes are not blocked in 866 since there is a path to node 889.

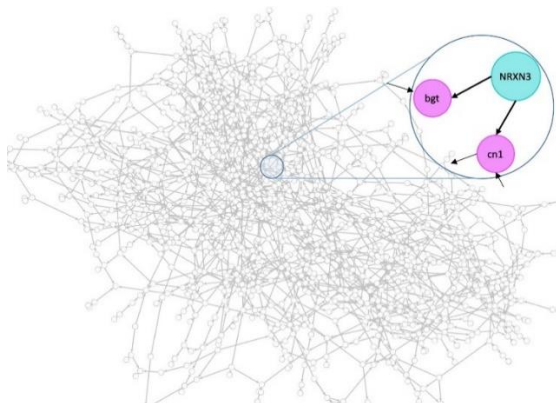

**Figure 5**. Extracted information from a causal network about principles governing the relationships. Identification of the underlying relationships through a causal network reveals a common cause (the blue node) of the two purple nodes associated with a disease. This kind of information is useful for experimental designs.

Through causal networks, we can see how the effect of an intervention spreads across the network, how the effect is blocked and which nodes/variables are influenced, see Figure 6 or (3)(22).

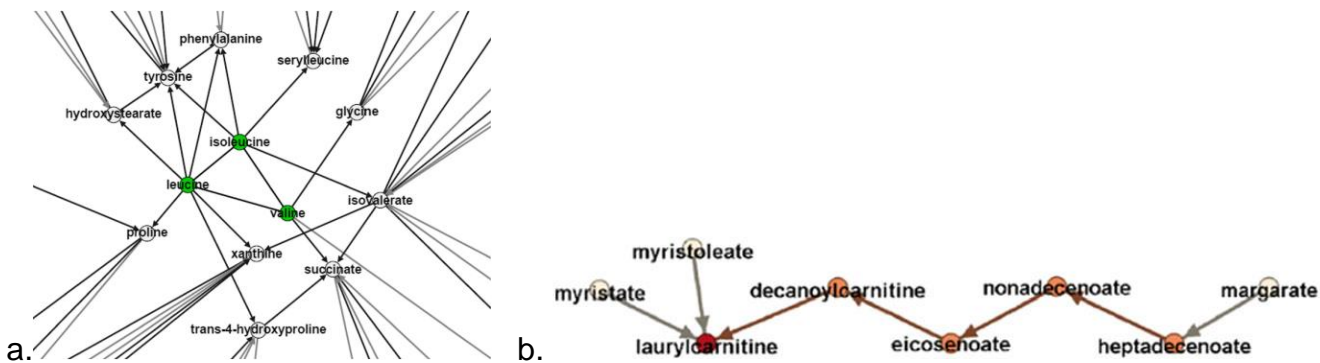

**Figure 6**. a. The effect of any intervention in leucine, isoleucine, or valine spreads across the system. b. The effect of margarate from one of its out degrees (arrows pointing away from margarate) spreads through 5 nodes/variables and is blocked in laurylcarnitine.

In a causal network, we can identify modules/subnetworks and the border of modules based on directions and causal effect size (22). Since the underlying structure between variables is identified (the causal network), we can graphically determine confounders (components in the system that act as confounders) and measure the causal effect size; we can also determine mediators to measure direct and indirect effect sizes. Multiple examples are provided in (10).

Since causal networks are identification of underlying relationships among components, we can assess the findings in genome wide association studies such as hypothesized pleiotropic actions using structural equation modeling, an application is discussed in (23). We could also identify components with high impact on disease endpoints as well as direct and indirect pathways to disease endpoints. These findings are not only a reduction of high number of components to a sufficient set but also identification of underlying processes to learn about disease (24)(25). For example see Figure 7, where genetic, metabolomics, and triglyceride levels are integrated systematically. The background depicts the metabolomics causal network identified by genetic information and then integrated with triglyceride levels. The focus is on underlying relationships among the four metabolites with direct paths to arachidonic acid, which has a direct path to triglycerides and the largest effect on it. To learn about triglycerides from metabolomics, we do not need to know about the levels of the four metabolites with direct effect on arachidonate since their effects on triglycerides is through arachidonate and arachidonate is sufficient. These novel findings have been validated clinically (26).

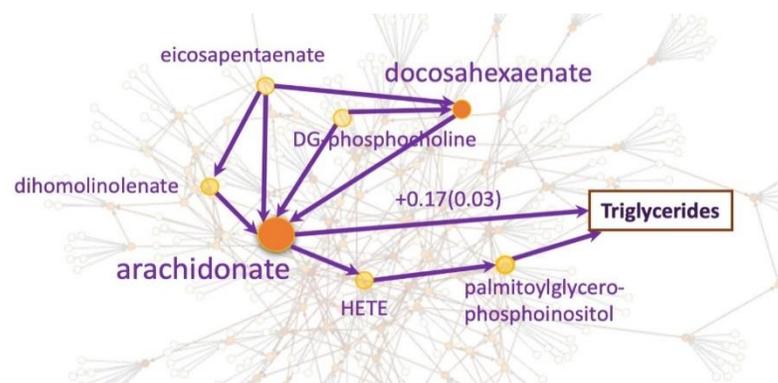

**Figure 7**. A systematic integration of genetic, metabolomics (as an intermediate molecular level), and triglyceride levels (as the outcome). This figure highlights underlying pathways from four metabolites with direct paths to arachidonic acid as well as pathways from arachidonate to triglycerides. DG-phosphocholine stands for Docosapentaenoyl-Glycerophos-phocholine and HETE for Hydroxyeicosatetraenoic acid.

Through a systematic integration of genetics, intermediate molecular levels/omics, and disease risk factors/disease end points, we can identify pathways across different biological levels of granularity. Figure 8 depicts indirect pathways from *FAM198B* and *C6orf25*, carriers of loss of function mutations, to triglycerides through metabolites HETE and palmitoylglycerophosphoinositol. We can see that the effect of these genes on triglycerides is not through arachidonate.

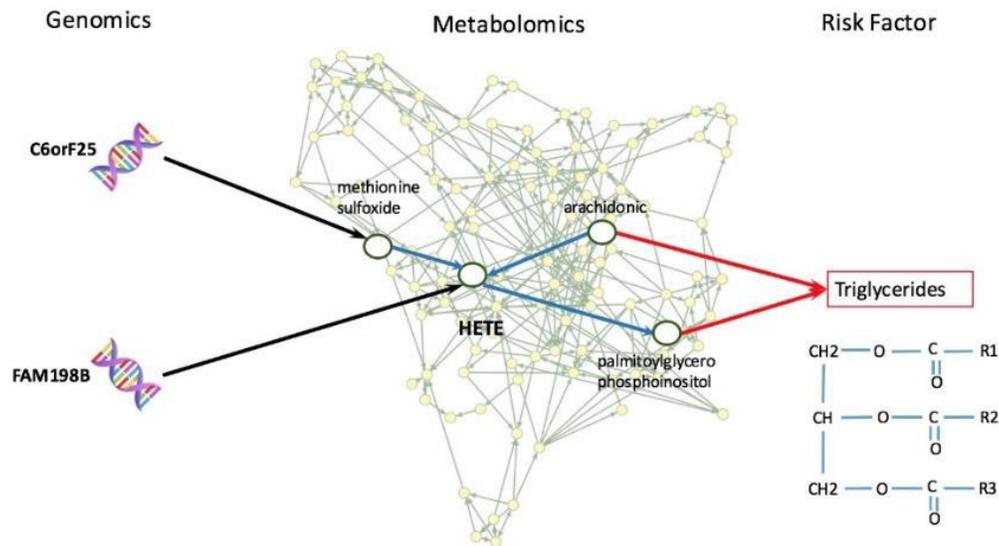

**Figure 8**. Identification of pathways across different biological levels of granularity. Indirect pathways from *FAM198B* and *C6orf25* to triglycerides through metabolites HETE and palmitoylglycerophosphoinositol and not through arachidonate.

**Discussion**

Complex mechanisms that employ multiple hierarchies cannot be understood by finding one causal factor. Parallel with evolving technologies to measure different granularities/omics with ever greater breadth, development of techniques to consider information from different granularities are essential to produce more descriptive models of the underlying biological processes. The tendency to focus on single variable and single omic analysis is a major limitation that prevents taking advantage of availability of comprehensive data profiles. While finding a causal relationship is one step further in association studies and we achieve some understanding in this way, progress is limited because it does not provide a complete context to interpret the findings.

Causal networks are the results of systematic integration of multi-omic data sets. Application of causal networks to systematically integrate different omics provides a path to mechanistic understanding including mechanistic understanding of the omic under study, finding the role of each component in disease processes as well as providing global insights that provide deep understanding for discovery, treatment development and further experimental studies.

Identification of causal networks are established in the principles of Mendelian randomization. Despite this fact, causal networks are not discussed in the field of Mendelian randomization including mediation analysis for direct and indirect causal effect measurements. The discussions in this field are mostly limited to two components (one exposure and one response) except in a recent case (27). There is a need for innovative approaches to make application of IVs possible for integrating and analyzing large scale omics systematically and uncovering underlying biological processes.

A difficulty for developing systems approaches is the requirement of diverse expertise and extensive collaboration. The van diagram in Figure 9 represents the intersection of three major fields of study: statistical modeling, computer science and biology. The intersection is an area that we can develop systems approaches to answer modern questions in biomedicine. Between each of these two fields there is a spectrum of other areas for research such as bioinformatics, biostatistics and bioengineering.

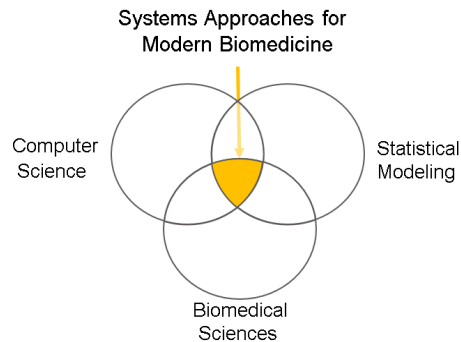

**Figure 9**. A van diagram representing diverse expertise and extensive collaboration as requirements to develop systems approach and answer modern questions in biomedicine.

We need to integrate techniques in network biology with the principles of Mendelian randomization and therefore, we need to modulate the application of IV for identification of large-scale casual networks. Then, we need to use statistical and machine learning techniques to extract information from the identified network. To move from information to knowledge, we need to interpret the findings with the science in biology. These are steps that we may need to take multiple times to learn from data, improve the techniques, and understand mechanisms.

Here, we discussed a systems approach to integrate multi omics for mechanistic understanding. Since the purpose of IV techniques is not finding genetic factors causal for phenotypes, we do not need to use natural IVs. Instead, we can generate IVs through extracting information from genetic variants using statistical techniques, such as principal component analysis. Since the genetic variants are abundant, we can generate multiple instrumental variables sufficient for identification of causal networks in large scale. This study explains the features of such an algorithm and how the underlying assumption is satisfied. In addition, this study briefly reviews the utilities of causal networks which are flourishing. The hope is that this work opens discussions in the field and as a result, new approaches emerge for systematic analysis of multi omics.

**Appendix**

**Causal network parameter**. Out-degree: The number of arrows pointing away from a node/variable is called the out-degree, which spreads information from the node to other nodes, Figure 4. In-degree: The number of arrows pointing into a node is called the in-degree and indicates influences from other nodes, Figure 4. The effect of the out-degree can be propagated serially, until reaching a blocking node, a node with zero out-degree. The effect blocking step is defined as the number of nodes influenced by one node in one path before the effect is blocked. Max. effect blocking step: The maximum number across paths leading out from a node is called the max. effect blocking step, Figure 6. Some nodes have high impact across the network due to having a high out-degree and a high number of max. effect blocking step and therefore, might be a good target for intervention. Nodes with high in-degree capture the effect of several other nodes in the system and therefore, might be a good target for prediction. A module is a subset of densely connected nodes sparsely connected with other nodes. We defined the module border based on effect sizes, connectivity, and direction of information flow (22). Directional information permits distinguishing confounders from covariates and measuring the causal effect size (29)(10).